\documentstyle[12pt]{article}
\newcommand{\re}{\par\hangindent=0.5cm\hangafter=1\noindent}

\newcommand{\lsim}{\raisebox{0.3mm}{\em $\, <$} \hspace{-3.3mm}
   \raisebox{-1.8mm}{\em $\sim \,$}}
\newcommand{\gsim}{\raisebox{0.3mm}{\em $\, >$} \hspace{-3.3mm}
   \raisebox{-1.8mm}{\em $\sim \,$}}

\newcommand{\bm}{\boldmath}

\newcommand{\bmi}{\mbox{\bm $i$}}

\newcommand{\bmv}{\mbox{\bm $v$}}
\newcommand{\bms}{\mbox{\bm $s$}}
\newcommand{\bmnabla}{\mbox{\bm $\nabla$}}
\newcommand{\bmF}{\mbox{\bm $F$}}

\begin{document}

\baselineskip 14pt

\begin{center}
{\large {\bf Hoyle--Lyttleton Accretion onto Accretion Disks}} \\
  
Jun {\sc Fukue} and Masayuki {\sc Ioroi} \\
{\it  Astronomical Institute, Osaka Kyoiku University,
           Asahigaoka, Kashiwara, Osaka 582-8582}\\
{\it E-mail(JF): fukue@cc.osaka-kyoiku.ac.jp}\\
\end{center}


\begin{center}
{\bf Abstract}
\end{center}

We investigate Hoyle--Lyttleton accretion
for the case where the central source is 
a luminous accretion disk.
In classical Hoyle-Lyttleton accretion onto a ``spherical'' source,
accretion takes place in an axially symmetric manner
around a so-called accretion axis.
In the spherical case
the accretion rate $\dot{M}$ is given as
$\dot{M}_{\rm HL} (1-\Gamma)^2$,
where
$\dot{M}_{\rm HL}$ is the accretion rate of
the classical Hoyle--Lyttleton accretion onto a non-luminous object
and $\Gamma$ the luminosity of the central object
normalized by the Eddington luminosity.
If the central object is a compact star
with a luminous accretion disk,
the radiation field becomes ``non-spherical''.
In such a case
the axial symmetry around the accretion axis breaks down;
the accretion radius $R_{\rm acc}$ generally depends on
an inclination angle $i$ between the accretion axis
and the symmetry axis of the disk
and the azimuthal angle $\varphi$ around the accretion axis.
Hence, 
the accretion rate $\dot{M}$,
which is obtained by integrating $R_{\rm acc}$ around $\varphi$,
depends on $i$. 
%
In the case of a pole-on accretion ($i=0$),
although the axisymmetry around $\varphi$ is retained,
the accretion radius becomes smaller than
that of the spherical case.
The accretion rate is therefore smaller than
that of the spherical case.
We found that the accretion rate is approximately expressed as
$\dot{M} \sim \dot{M}_{\rm HL} (1-\Gamma)(1-2\Gamma)$.
In the case of an edge-on accretion ($i=90^{\circ}$),
the accretion radius depends strongly on $\varphi$
and is somewhat larger than that of the spherical case.
Depending on the central luminosity $\Gamma$,
the shape of the accretion cross-section varies from
a circle ($\Gamma=0$), an ellipse, 
a hollow ellipse ($\Gamma \sim 0.5$), and
a twin lobe ($\Gamma \gsim 0.65$).
The accretion rate is larger than that of the spherical case
and approximately expressed as
$\dot{M} \sim \dot{M}_{\rm HL} (1-\Gamma)$ for $\Gamma \leq 0.65$
and
$\dot{M} \sim \dot{M}_{\rm HL} (2-\Gamma)^2/5$ for $\Gamma \geq 0.65$.
Once the accretion disk forms and
the anisotropic radiation fields are produced
around the central object,
the accretion plane will be maintained automatically
(the direction of jets associated with the disk is also maintained).
Thus, the anisotropic radiation field of accretion disks
drastically changes the accretion nature, that
gives a clue to the formation of accretion disks
around an isolated black hole.
Hoyle--Lyttleton type accretion onto the accretion disk
may take place in various astrophysical situations,
including the Galactic Center X-ray source 1E~1740.7$-$2942.



\newpage

\begin{flushleft}
{\bf 1. Introduction}
\end{flushleft}

Accretion onto a gravitating body
is one of the most important phenomena in recent astrophysics,
since it produces and radiates a remarkable energy
via the release of the gravitational energy (e.g., Kato et al. 1998).
If the central object is a compact star,
a significant fraction of the rest mass energy of
the accreting gas is liberated.
Furthermore, an energetic mass outflow from the accretion system
often takes place.
Thus, the central engine of the accretion system
exerts a strong influence on the environments
through intense radiation and mass outflow.

If there is an isolated object in interstellar space,
the gravitational field of the object
attracts gas particles.
If there exists a relative velocity
between the object and the gas,
the gas particles move in hyperbolic orbits,
which intersect downstream of the point mass 
on the axis of symmetry (the accretion axis). 
The intersecting particles are allowed to collide 
and coalesce on the accretion axis. 
Particles colliding sufficiently close to the point mass lose 
enough of their kinetic energy to be trapped by the central object. 
This is the classical picture of 
Hoyle--Lyttleton accretion 
(Hoyle, Lyttleton 1939; Bondi, Hoyle 1944).

Let us suppose 
a graviating body with mass $M$ and luminosity $L$,
placed in a gas flow
with velocity $v_{\infty}$ and density $\rho_{\infty}$
at infinity.
If the velocity $v_{\infty}$ is much larger 
than the sound speed of the gas at infinity,
the effect of pressure can be neglected,
except at the very center.
The Hoyle--Lyttleton accretion radius $R_{\rm HL}$, 
where the effect of radiation pressure is ignored, is
\begin{eqnarray}
   R_{\rm HL} &=& \frac{2GM}{v_{\infty}^2}
\nonumber \\
   &=& 2.65 \times 10^{15} 
            \frac{M}{10M_{\odot}}
            \left( \frac{v_\infty}{10{\rm ~km~s}^{-1}} \right)^{-2}
            {\rm ~cm}.
\end{eqnarray}
Here, we bear in mind a stellar mass black hole
moving slowly in the interstellar space.
In addition, the relevant timescale of
the classical Hoyle--Lyttleton accretion is
\begin{eqnarray}
   t_{\rm HL} &=& \frac{R_{\rm HL}}{v_\infty}
\nonumber \\
   &=& 84.0 
            \frac{M}{10M_{\odot}}
            \left( \frac{v_\infty}{10{\rm ~km~s}^{-1}} \right)^{-3}
            {\rm ~yr}.
\label{tHL}
\end{eqnarray}

Inside the accretion radius
the potential energy exceeds the kinetic energy
and the gas is trapped by the gravitating object.
Thus, the mass accretion rate ${\dot M}_{\rm HL}$
onto the gravitating object is
\begin{eqnarray}
   {\dot M}_{\rm HL} &=& \pi R_{\rm HL}^2 \rho_{\infty} v_{\infty}
           = \frac{4\pi\rho_{\infty}G^2 M^2}{v_{\infty}^3}
\nonumber \\
   &=& 3.70 \times 10^{18}
            \left( \frac{M}{10M_{\odot}} \right)^2
            \frac{n_\infty}{10^5 {\rm ~cm}^{-3}}
            \left( \frac{v_\infty}{10{\rm ~km~s}^{-1}} \right)^{-3}
            {\rm ~g~s}^{-1},
\end{eqnarray}
where $n_\infty$ is the number density at infinity.
In addition, the growth time of the central object is
\begin{eqnarray}
   t_{\rm growth} &=& \frac{M}{\dot{M}_{\rm HL}}
\nonumber \\
   &=& 1.70 \times 10^8 
            \left( \frac{M}{10M_{\odot}} \right)^{-1}
            \left( \frac{n_\infty}{10^5 {\rm ~cm}^{-3}} \right)^{-1}
            \left( \frac{v_\infty}{10{\rm ~km~s}^{-1}} \right)^{3}
            {\rm ~yr}.
\label{tgrowth}
\end{eqnarray}
If we know the mass accretion rate, 
we can estimate the (accretion) luminosity of the radiation 
emitted from the compact object. 
Therefore, knowledge about the accretion radius is important 
to compare theories with observations.

The effect of gas pressure on Hoyle--Lyttleton accretion
has been investigated by many researchers
(Bondi 1952; Hunt 1971; Shima et al. 1985; Fryxell et al. 1987;
Fryxell, Taam 1988; Taam, Fryxell 1989; Ho et al. 1989; 
Matsuda et al. 1991;
Ruffert, Arnett 1944; Ruffert 1994, 1996;
Font, Ib\'a\~nez 1998a, b).
The effect of radiation pressure, however,
 has not been taken into account well (e.g., Taam et al. 1991).
Due to radiation pressure,
the accretion radius and the accretion rate
are significantly reduced from the Hoyle--Lyttleton estimate.
That is,
the accretion radius modified by the radiation pressure becomes
\begin{equation}
   R_{\rm HL}^{\rm rad}=\frac{2GM(1-\Gamma)}{v_{\infty}^2}
   =R_{\rm HL}(1-\Gamma),
\label{eqn:bondi}
\end{equation}
when the gas is fully ionized and transparent.
Here, $\Gamma$ is the normalized luminosity,
which is defined by the ratio of the luminosity $L$ 
of the gravitating object to the Eddington luminosity 
$L_{\rm E}$ ($=4 \pi cGMm_{\rm p}/\sigma_{\rm T}$):
\begin{equation}
       \Gamma = \frac{L}{L_{\rm E}},
\end{equation}
where $m_{\rm p}$ is the proton mass
and $\sigma_{\rm T}$ the Thomson scattering cross-section.
In this case with radiation pressure,
$\dot{M}=\dot{M}_{\rm HL}(1-\Gamma)^2$.

Recently,
the effect of radiation drag (Compton drag)
in Hoyle--Lyttleton accretion around a luminous source
was also considered by Nio et al. (1998).
In the ionized gas the radiation-drag force becomes important
if the velocity of the gas is comparable to the light speed
and the luminosity of the central object is 
comparable to the Eddington luminosity.
Radiation drag is also important
if the specific cross-section --- ``opacity'' ---
of particles is sufficiently large like, e.g., dust.
Nio et al. (1998) found that
the accretion radius decreases due to radiation pressure,
while it increases due to radiation drag.
In addition, the accretion radius including radiation drag
becomes larger as the incident velocity becomes fast.
In the limit of small $v_\infty$,
the above equation (\ref{eqn:bondi}) is modified to
$R_{\rm HL}^{\rm rad}/R_{\rm HL} = 1 - \Gamma (1-2v_\infty/c)$.

In all of these studies
the central sources have been supposed to be ``spherical''
in the sense that
both of the gravitational and radiation fields
have a spherical symmetry.
In actual accretion systems
like the Galactic Center X-ray source 1E~1740.7$-$2942, however,
as a consequence of mass accretion,
an accretion disk often forms
around the central gravitating body.
Hence, in such a case
the radiation field is not spherical but ``anisotropic''.
Namely, the radiative flux $F$ at a distance $R$ from the center
depends on a polar angle $\theta$ as
\begin{equation}
   F= \frac{L}{2\pi R^2} \cos \theta,
\label{diskF}
\end{equation}
where $L$ is the total luminosity of the disk.
The accretion radius and therefore the mass accretion rate
will be changed due to the violation of the spherical symmetry.
Indeed, the anisotropic radiation field
drastically changes the nature of accretion.
The purpose of this paper is to examine
the Hoyle--Lyttleton accretion onto an accretion disk.

In the next section 
we describe situations and basic equations.
Numerical results are presented in section 3.
The canonical luminosity is derived in section 4.
Astrophysical implications and applications are discussed in section 5.
The final section is devoted to concluding remarks.
The physical quantities of the X-ray source 1E~1740.7$-$2942
are deduced in the appendix.

\begin{flushleft}
{\bf 2. Situations and Basic Equations}
\end{flushleft}

\begin{flushleft}
{\it 2.1. Radiation Fields}
\end{flushleft}

Let us consider 
a point mass with mass $M$ and
a surrounding accretion disk with luminosity $L$ (figure 1).
The system is assumed to be immersed in uniform gas flow
with velocity $v_\infty$ and density $\rho_\infty$.
We also assume that the gas is fully ionized and transparent 
to radiation from the object.
The central accretion disk is generally
inclined to the gas flow
(the inclination angle between
the accretion axis and the symmetry axis of the disk
is $i$).
We adopt cylindrical coordinates $(r, \varphi, z)$,
where the $z$-axis ({\it accretion axis}) is in
the direction of the upstream gas particles (see figure 1).
Moreover, the coordinates of a particle coming from infinity
are $(x, y, z)$ in Cartesian coordinates and
$(R, \theta, \varphi)$ in spherical coordinates
($R=\sqrt{r^2+z^2})$.
In this system
we first derive the radiative flux vector
at a point P, where a particle is located.

\begin{center}
------------ \\
figure 1 \\
------------ \\
\end{center}

In Cartesian coordinates,
the unit vector $\bmi$
in the direction of the symmetry axis of the disk,
and the unit vector $\bms$ 
in the direction of a particle are respectively
\begin{eqnarray}
   \bmi &=& \left( \sin i, 0, \cos i \right),
\\
   \bms &=& \left( \frac{r}{R}\cos\varphi,
                   \frac{r}{R}\sin\varphi, \frac{z}{R} \right).
\end{eqnarray}
Hence, the angle $\psi$ between
the symmetry axis $\bmi$ and the particle direction $\bms$ is
given by
\begin{equation}
   \cos \psi = {\bmi} \cdot {\bms}
             = \frac{1}{R} 
               \left( r\cos \varphi \sin i + z \cos i \right).
\end{equation}

The radiative flux $F_R$ in the $R$-direction
at a point (distance $R$) is expressed as
\begin{equation}
   F_R = \frac{L \cos \psi}{2\pi R^2}.
\label{FR}
\end{equation}
Consequently, the radial component $F_r$ and
the vertical component $F_z$ in cylindrical coordinates are,
respectively,
{
\setcounter{enumi}{\value{equation}}
\addtocounter{enumi}{1}
\setcounter{equation}{0}
\renewcommand{\theequation}{\theenumi\alph{equation}}
\begin{eqnarray}
   F_r &=& F_R \frac{r}{R},
\\
   F_z &=& F_R \frac{z}{R}.
\end{eqnarray}
\setcounter{equation}{\value{enumi}}
}

It should be noted that
if the distance from the center is of the order of the disk size,
the radiation fields of the disk
becomes much more complicated (cf. Tajima, Fukue 1998)
and the above formula becomes inadequate.

\begin{flushleft}
{\it 2.2. Basic Equations}
\end{flushleft}

The motion of the gas particle in radiation fields
is described by
\begin{equation}
\frac{d\bmv}{dt}
     =-\bmnabla \phi + \frac{\sigma_{\rm T}}{mc}
       (\bmF-E \bmv -{\bf P} \otimes \bmv)
\label{vector}
\end{equation}
upto the first order of $v/c$,
where $\bmv$ is the particle velocity,
$\phi$ the gravitational potential,
$m$ the particle mass (proton mass for the normal plasma),
$E$ the radiation energy density,
$\bmF$ the radiative flux vector,
and {\bf P} the radiation stress tensor
(Hsieh, Spiegel 1976; Fukue et al. 1985;
see also Kato et al. 1998).

To examine the effect of radiation drag,
Nio et al. (1998) considered the radiation energy and
the radiation stress tensor as well as the radiative flux.
In the present study
we only consider the radiative flux
to avoid the complication
and to focus our attention on the anisotropic feature
of the radiation field.
Thus, the equation of motion
in cylindlical coordinates $(r, \varphi, z)$
is expressed as
{
\setcounter{enumi}{\value{equation}}
\addtocounter{enumi}{1}
\setcounter{equation}{0}
\renewcommand{\theequation}{\theenumi\alph{equation}}
\begin{eqnarray}
    \frac{\displaystyle dv_r}{\displaystyle dt} 
    &=& -\frac{\displaystyle GMr}{\displaystyle R^3}
        +\frac{\displaystyle \sigma_{\rm T}}{\displaystyle mc}
         F_R \frac{\displaystyle r}{\displaystyle R}, 
\\
    \frac{\displaystyle dv_z}{\displaystyle dt} 
    &=& -\frac{\displaystyle GMz}{\displaystyle R^3}
        +\frac{\displaystyle \sigma_{\rm T}}{\displaystyle mc}
         F_R \frac{\displaystyle z}{\displaystyle R}.  
\label{eqn:basiceq0}
\end{eqnarray} 
\setcounter{equation}{\value{enumi}}}
Using equation (\ref{FR}),
the above equation of motion becomes
{
\setcounter{enumi}{\value{equation}}
\addtocounter{enumi}{1}
\setcounter{equation}{0}
\renewcommand{\theequation}{\theenumi\alph{equation}}
\begin{eqnarray}
    \frac{\displaystyle dv_r}{\displaystyle dt} 
    &=&-\frac{\displaystyle GMr}
             {\displaystyle R^3}(1-\Gamma_{\rm eff}),
\\
    \frac{\displaystyle dv_z}{\displaystyle dt} 
    &=&-\frac{\displaystyle GMz}
             {\displaystyle R^3}(1-\Gamma_{\rm eff}),
\end{eqnarray}
\setcounter{equation}{\value{enumi}}}
where $\Gamma_{\rm eff}$ is the {\it effective} normalized luminosity
defined by
\begin{eqnarray}
   \Gamma_{\rm eff} &\equiv& 2 \Gamma \cos\psi
\nonumber \\
             &=& 2\frac{L}{L_{\rm E}}
                 \frac{ r\cos \varphi \sin i + z \cos i }
                      {R}.
\end{eqnarray}
Measuring the length and velocity in units of
$R_{\rm HL}$ and $v_\infty$, respectively,
and using the relevant timescale $t_{\rm HL}$,
the equation of motion is finally rewritten in the dimensionless form:
{
\setcounter{enumi}{\value{equation}}
\addtocounter{enumi}{1}
\setcounter{equation}{0}
\renewcommand{\theequation}{\theenumi\alph{equation}}
\begin{eqnarray}
    \frac{\displaystyle d\hat{v}_r}{\displaystyle d\hat{t}} 
    &=&-\frac{\displaystyle \hat{r}}
             {\displaystyle 2\hat{R}^3}(1-\Gamma_{\rm eff}),
\\
    \frac{\displaystyle d\hat{v}_z}{\displaystyle d\hat{t}} 
    &=&-\frac{\displaystyle \hat{z}}
             {\displaystyle 2\hat{R}^3}(1-\Gamma_{\rm eff}).
\label{eqn:basiceq}
\end{eqnarray}
\setcounter{equation}{\value{enumi}}}

For various combinations of initial positions $(r, \varphi)$
of particles at sufficiently large $z$,
we numerically integrate equation (17) 
using the forth-order Runge--Kutta--Gill method
and calculate trajectories of particles.
We estimate the accretion radius by the same procedure
as that of Hoyle and Lyttleton (1939).
That is, according to parameters,
particles intersect the accretion axis
at the downstream of the axis.
If the $z$-velocity is less than the escape velocity
at that point ($z=z_0$),
defined by 
$\hat{v}_{\rm esc}=
\sqrt{(1-\Gamma_{\rm eff})/\hat{z}}
\left|_{r=0, z=z0}\right.$, 
these particles are judged to be
captured and accreted on the gravitating body.

The parameters of the present problem are
the normalized luminosity $\Gamma$,
the incident velocity $v_\infty$
(although $v_\infty$ is just the unit
and therefore renormalized), and
the disk inclination angle $i$.
In the present study
we only examine two extreme cases;
a pole-on accretion ($i=0$) and an edge-on one ($i=90^{\circ}$).
Other general cases with arbitrary $i$ lie
between these two cases.

\begin{flushleft}
{\bf 3. Accretion Radius and Accretion Rate}
\end{flushleft}

Figure 2 shows the accretion radius $R_{\rm acc}$
normalized by the Hoyle--Lyttleton accretion radius $R_{\rm HL}$
($=2GM/v_{\infty}^2$)
as a function of the normalized luminosity $\Gamma$ ($=L/L_{\rm E}$).

\begin{center}
------------ \\
figure 2 \\
------------ \\
\end{center}

The dotted straight line is for the case of a spherical source,
that is proportional to $(1-\Gamma)$,
as shown in equation (\ref{eqn:bondi}).
The dashed curve is the pole-on accretion case ($i=0$),
where the accretion radius is independent of $\varphi$.
As already stated in equation (\ref{diskF}),
the disk radiation field is not spherical but
orients towards the polar direction.
Hence, in the pole-on accretion of $i=0$,
the central source (disk) is effectively luminous,
compared with the spherical case.
As a result,
the accretion radius is remarkably smaller than
that of the spherical case, as seen in figure 2.

The solid curves are the edge-on accretion case ($i=90^{\circ}$),
where the accretion radius strongly depends on $\varphi$.
In this case
the situation is much more complicated,
because the radiation strength varies along the particle path
and depends on $\varphi$.
Roughly speaking, however,
in the edge-on accretion of $i=90^{\circ}$,
the disk is effectively less luminous,
and the accretion radius is larger than
that of the spherical case.

In figure 3
the shapes of cross section in the case of $i=90^{\circ}$
are shown for various $\Gamma$.
Since incoming particles in the disk plane ($\varphi=90^{\circ}$)
receive no radiative flux,
the accretion radius in the disk plane
is just the classical Hoyle--Lyttleton one $R_{\rm HL}$.
On the other hand,
particles travelling over the disk are so influenced
by the disk radiation field that
the accretion radius becomes smaller than $R_{\rm HL}$.
Moreover, particles passing through just above the pole
($\varphi=0$)
cannot accrete for $\Gamma \gsim 0.65$.
As a result,
depending on the normalized luminosity $\Gamma$,
the shape of the accretion cross-section varies
from a circle ($\Gamma=0$),
an ellipse compressed perpendicular to the disk plane, 
a hollow ellipse ($\Gamma \sim 0.5$),
and a twin lobe in the disk plane ($\Gamma \gsim 0.65$).

\begin{center}
------------ \\
figure 3 \\
------------ \\
\end{center}

Since in the present situation
the accretion radius $R_{\rm acc}$ generally depends on $\varphi$,
the cross sectional area of accretion is more convenient
to discuss the accretion nature.
In addition, the cross sectional area $A_{\rm acc}$ of accretion
is directly related to the accretion rate by
$\dot{M}= A_{\rm acc} \rho_\infty v_\infty$.

In figure 4 we show 
the cross sectional area $A_{\rm acc}$,
normalized by that of
the Hoyle--Lyttleton accretion, $\pi R_{\rm HL}^2$,
as a function of the normalized luminosity $\Gamma$
(this also expresses $\dot{M}/\dot{M}_{\rm HL}$).
As expected from the accretion radius calculated above,
the accretion rate in the pole-on case is smaller than
that of the spherical case.
The accretion rate in the edge-on case, on the other hand,
is larger than that of the spherical case.
It should be noted that in the edge-on case
accretion is possible even if $\Gamma$ is unity!

\begin{center}
------------ \\
figure 4 \\
------------ \\
\end{center}

Finally, from these numerical results,
the cross sectional area $A_{\rm acc}$
is approximately expressed as
\begin{equation}
   A_{\rm acc} = \pi R_{\rm HL}^2 f(\Gamma, i),
\end{equation}
where
\begin{equation}
   f\left(\Gamma, i\right) = \left\{
     \begin{array}{ll}
       (1-\Gamma)^2 & ~~~{\rm for~a~spherical~case} \\
       (1-\Gamma)(1-2\Gamma) & ~~~{\rm for~}i=0 \\
       (1-\Gamma) & ~~~{\rm for~}i=90^{\circ}, \Gamma \leq 0.65 \\
       (2-\Gamma)^2/5 & ~~~{\rm for~}i=90^{\circ}, \Gamma \geq 0.65.
     \end{array}
   \right.
\label{f}
\end{equation}

\begin{flushleft}
{\bf 4. Canonical Luminosity}
\end{flushleft}

In the previous section,
we treat the central luminosity $L$
and the accretion rate $\dot{M}$, seperately.
In the accretion phenomena, however,
these quantities are closely connected via the accretion process.
That is,
if the Hoyle--Lyttleton accretion takes place
on an isolated compact star,
the accretion rate $\dot{M}$ determines
the central (accretion) luminosity $L$.
We here estimate the {\it canonical} luminosity $L_{\rm can}$
(or equivalently $\Gamma_{\rm can}$) under such a situation.
For simplicity,
we assume a steady state of accretion,
although there may exist a time lag
for the gas to accrete along the accretion axis toward the center
(e.g., Taam et al. 1991).

As shown in the previous section,
the cross sectional area $A_{\rm acc}$ is given by
$A_{\rm acc}=\pi R_{\rm HL}^2 f(\Gamma, i)$,
and therefore,
the accretion rate $\dot{M}$ 
($=A_{\rm acc}\rho_\infty v_\infty$) is given by
\begin{equation}
   \frac{\dot{M}}{\dot{M}_{\rm HL}} = f(\Gamma, i)
\label{dM1}
\end{equation}
for the present situation 
(see figure 4 for a representation of the function $f$).
On the other hand,
the accretion luminosity $L$ follows the mass accretion rate $\dot{M}$
via $L=\eta \dot{M}c^2$
(and therefore, $\Gamma=\eta \dot{M}c^2/L_{\rm E}$),
where $\eta$ is the efficiency of the release
of the accretion energy.
Introducing the critical accretion rate $\dot{M}_{\rm E}$ by
$\dot{M}_{\rm E} \equiv L_{\rm E}/(\eta c^2)$,
we finally have
\begin{equation}
   \frac{\dot{M}}{\dot{M}_{\rm HL}} 
   = \frac{\dot{M}_{\rm E}}{\dot{M}_{\rm HL}}\Gamma
   = \frac{1}{\dot{m}_{\rm HL}}\Gamma,
\label{dM2}
\end{equation}
where $\dot{m}_{\rm HL}$ is the dimensionless parameter
of the system defined by
\begin{eqnarray}
   \dot{m}_{\rm HL} &\equiv& \frac{\dot{M}_{\rm HL}}{\dot{M}_{\rm E}}
   = \frac{\eta cGM\sigma_{\rm T}\rho_\infty}{m_{\rm p}v_\infty^3}
\nonumber \\
   &=& 0.2648
       \frac{\eta}{0.1}
            \frac{M}{10M_{\odot}}
            \frac{n_\infty}{10^5 {\rm ~cm}^{-3}}
            \left( \frac{v_\infty}{10{\rm ~km~s}^{-1}} \right)^{-3}.
\label{dmHL}
\end{eqnarray}
This parameter $\dot{m}_{\rm HL}$ is just
$\alpha$ in Taam et al. (1991),
where the spherical case was considered.

For a given configuration of the system (source type or inclination)
and a given parameter $\dot{m}_{\rm HL}$,
these two relations (\ref{dM1}) and (\ref{dM2})
have an intersection point on the $(\Gamma, \dot{M})$-plane
(cf. figure 4). 
Moreover, this solution is stable
in the sense that the increase of the accretion rate
leads to the increase of the luminosity,
then the accretion rate decreases.
This intersection (solution) thus gives
the {\it canonical luminosity} $L_{\rm can}$
(or $\Gamma_{\rm can}$)
and the {\it canonical accretion rate} $\dot{M}_{\rm can}$.
In figure 5
the canonical luminosity are shown as a function of $\dot{m}_{\rm HL}$. 
The canonical accretion rate is obtained by
$\dot{M}_{\rm can}/\dot{M}_{\rm E} 
= L_{\rm can}/L_{\rm E} = \Gamma_{\rm can}$
or by
$\dot{M}_{\rm can}/\dot{M}_{\rm HL}
=\Gamma_{\rm can}/\dot{m}_{\rm HL}$.

\begin{center}
------------ \\
figure 5 \\
------------ \\
\end{center}

As seen in figure 5,
the canonical luminosity $\Gamma_{\rm can}$
increases with $\dot{m}_{\rm HL}$.
Compared with the spherical case (dotted curve),
in the pole-on case (dashed one)
the canonical luminosity is small,
whereas it is large
in the edge-on case (solid one).
It is easy to obtain $\Gamma_{\rm can}$ algebraically
as a function of $\dot{m}_{\rm HL}$:
\begin{equation}
   \Gamma_{\rm can} = \left\{
     \begin{array}{ll}
       \frac
            {\displaystyle 2\dot{m}_{\rm HL}+1
              -\sqrt{4\dot{m}_{\rm HL}+1} } 
            {\displaystyle 2\dot{m}_{\rm HL}}
       & ~~~{\rm for~a~spherical~case} \\
       \frac
            {\displaystyle 3\dot{m}_{\rm HL}+1
              -\sqrt{\dot{m}_{\rm HL}^2+6\dot{m}_{\rm HL}+1} }
            {\displaystyle 4\dot{m}_{\rm HL}}
       & ~~~{\rm for~}i=0 \\
       \frac{\displaystyle \dot{m}_{\rm HL}}
            {\displaystyle \dot{m}_{\rm HL}+1}
       & ~~~{\rm for~}i=90^{\circ}, \Gamma \leq 0.65 \\
       \frac
            {\displaystyle 4\dot{m}_{\rm HL}+5
              -\sqrt{40\dot{m}_{\rm HL}+25} }
            {\displaystyle 2\dot{m}_{\rm HL}}
       & ~~~{\rm for~}i=90^{\circ}, \Gamma \geq 0.65.
     \end{array}
   \right.
\end{equation}

As accretion proceeds,
the mass of the central object gradually increases,
while producing the enormous luminosity.
This growth timescale $t_{\rm growth}$ is roughly
given by equation (\ref{tgrowth}).
Since the parameter $\dot{m}_{\rm HL}$ increases with mass $M$
and the canonical luminosity $\Gamma_{\rm can}$
also increases with $\dot{m}_{\rm HL}$,
the canonical luminosity $\Gamma_{\rm can}$ 
increases as the mass increases.
In other words,
after sufficiently large time,
$\Gamma_{\rm can}$ approaches unity.

\begin{flushleft}
{\bf 5. Discussion}
\end{flushleft}

In this section
we shall briefly discuss
the astrophysical implications and applications of 
the present Hoyle--Lyttleton accretion onto accretion disks.

\begin{flushleft}
{\it 5.1. Maintenance of the Disk Plane}
\end{flushleft}

As shown in the previous sections,
the anisotropic radiation fields of accretion disks
may drastically change the accretion nature of 
the classical Hoyle--Lyttleton accretion.

First, let us suppose the edge-on accretion ($i=90^{\circ}$).
Because of the anisotropy of radiation fields,
the accretion radius over the disk poleward becomes small,
compared with that in the disk plane (see figure 3).
In other words, accretion takes place mainly
in the disk plane.
This property,
that {\it the prefered accretion plane coincides with the disk plane},
thus continues to maintain the accretion disk.
Naively speaking, of course,
the net angular momentum of the accreting gas 
in the disk plane is zero.
Hence, the spin direction of the disk
may change on a timescale of $t_{\rm HL}$.
However, the disk always forms
with a fixed axis of symmetry.
These properties are essentially equal in general cases
with arbitrary inclination angle,
except for the pole-on accretion case.

Now, let us investigate the longterm behavior of the accretion.
As is seen from figure 4,
the accretion rate of the edge-on case is larger than other cases;
the edge-on accretion is the most effective.
Hence, if the disk was initially inclined to the gas flow
with arbitrary inclination angle,
the inclination angle would gradually increase, 
as accretion proceeds.
{\it Ultimately,
the disk configuration will settle down
into a state, where the disk axis becomes
perpendicular to the accretion axis}.
The final outcome would be edge-on accretion.

Finally, let us look at
the origin and maintenance of accretion disks,
which form around the gravitating object
during the Hoyle--Lyttleton accretion process.
At the first stage,
where only the gravitating body exists in the interstellar space,
there is no prefered direction of accretion.
So,
the classical Hoyle--Lyttleton accretion takes place
symmetrically around the accretion axis.
After the Hoyle--Lyttleton accretion proceeds,
some possible fluctuations in the gas density or velocity
break the symmetry of accretion around the accretion axis.
{\it Due to this spontaneous breaking of symmetry,
an accretion disk forms and a prefered direction is chosen}.
Once the disk forms,
the prefered direction and the disk plane
will be fixed and maintained, as discussed above.
If astrophysical jets emanate from the disk,
the direction of jets is also constant.

\begin{flushleft}
{\it 5.2. Application to 1E~1740.7$-$2942}
\end{flushleft}

We apply the present model
to the Galactic Center X-ray source 1E~1740.7$-$2942.

Electron-positron pair annihilation lines
 from the Galactic Center region
have been detected since the 1970's 
(e.g., Johnson et al. 1972; see Morris 1989 for a review).
The intensity of annihilation lines is about 
$10^{-3}$photons s${}^{-1}$ cm${}^{-2}$,
and this yields a positron annihilation rate of $10^{43}$ s${}^{-1}$
with the distance to the Galactic Center
(Johnson, Haymes 1973; Leventhal et al. 1978).
The lack of significant redshift and the narrowness of the line width
suggest that the pair annihilation takes place
in a relatively cold gas.
Recent observations by the GRANAT satellite identified
the pair annihilation source to be 
the hard X-ray source 1E~1740.7$-$2942 near the Galactic Center
(Sunyaev et al. 1991; Bouchet et al. 1991).
Subsequently, this X-ray source 1E~1740.7$-$2942 
was found to be embedded
in the molecular cloud which belongs to the inner molecular layer
of the Galactic Center region
(Mirabel et al. 1991; see also Ramaty et al. 1992).
This cloud has a size of $\sim$ 3pc and a mean density of 
$\sim 5\times 10^4 {\rm cm}^{-3}$,
therefore, it has a mass of $\sim 5 \times 10^4 M_{\odot}$.
Finally, observation with the Very Large Array
(Mirabel et al. 1992) showed
a double-sided radio jet from this X-ray source 1E~1740.7$-$2942.
The projected length of this radio jet is $\sim$ 1pc.

Several observational facts suggest the following picture 
for the X-ray source 1E~1740.7$-$2942 and the radio jet
(Mirabel et al. 1991, 1992).
The X-ray source 1E~1740.7$-$2942 may be 
a stellar mass black hole without companion
--- an {\it isolated black hole}
(the X-ray spectrum of 1E~1740.7$-$2942 resembles that of Cyg~X-1).
 From a molecular cloud, 
the interstellar gas accretes onto this black hole
via the Hoyle--Lyttleton process,
an accretion disk forms around the black hole,
and electron-positron pairs are created 
in the vicinity of the black hole.
The electron-positron pair plasma is ejected
perpendicularly to the disk plane (cf. Tajima, Fukue 1998), 
and then injected 
into the molecular cloud in the form of a double-sided jet
(the radio emission from jets is supposed to be synchrotron radiation
from electrons and positrons).
Electron-positron pairs travel at high velocities for a few years
and then they are slow down and annihilated
in the high-density cold molecular cloud
(the annihilation lines are narrow).

Now, using the observational facts described above and
physical quantities deduced from observations (in the appendix),
we shall estimate the plausible accretion rate
and the intrinsic luminosity of 1E~1740.7$-$2942.

We first evaluate the relative velocity $v_\infty$ at infinity
by geometrical reasoning and by looking at activity duration.
As already discussed,
in Hoyle--Lyttleton accretion onto the accretion disk,
the symmetry axis of the disk generally tends to lie
in the plane perpendicular to the flow direction.
If this is true for the case of 1E~1740.7$-$2942,
the black hole moves  perpendicular to the line-of-sight,
and the transverse velocity is of the order of $v_\infty$.
VLA observations show that
the jets in 1E~1740.7$-$2942 are rather straight
and no prominent bending is seen.
This geometrical properties mean that 
the ratio of the transverse velocity to the jet velocity
is less than $\sim 10^4$.
Since the jet velocity is estimated as $0.3c$ in the appendix,
the transverse velocity and therefore $v_\infty$ is about
$v_\infty \lsim 10 {\rm ~km~s}^{-1}$.
On the other hand, from
the projection length of jets ($\sim$ 1~pc) 
and the jet velocity ($0.3c$),
the central activity producing jets lasts
at least 10~yrs.
Due to the possible irregularities,
the direction of the disk rotation changes intermittently.
Indeed, the electron-positron annihilation features
disappear in a timescale of a few years.
These dynamical variations occur in the timescale of $t_{\rm HL}$
given in equation (\ref{tHL}).
Hence, the activity timescale again means that 
$v_\infty \sim 10 {\rm ~km~s}^{-1}$
as long as $M \sim 10M_{\odot}$.

Thus, for 1E~1740.7$-$2942, since
the mass $M$ is reasonably assumed to be $10M_{\odot}$,
a mean density of the ambient cloud is observed to be
$\sim 5\times 10^4 {\rm cm}^{-3}$, and
the velocity $v_\infty$ is derived to be $\lsim 10 {\rm ~km~s}^{-1}$,
we can estimate the parameter $\dot{m}_{\rm HL}$ as 
$\dot{m}_{\rm HL} \gsim 0.1$.

Furthermore, if the canonical luminosity is established in
1E~1740.7$-$2942,
then the intrinsic luminosity is $\Gamma \gsim 0.1$
(or $L \gsim 1.25 \times 10^{38} {\rm ~erg~s}^{-1}$ 
for $M=10M_{\odot}$).
The observational luminosity, on the other hand,
is about $3 \times 10^{37} {\rm ~erg~s}^{-1}$
(or normalized luminosity is 0.02).
Since the observational luminosity is the projected one and
equal to $2\Gamma \cos i$,
the inclination angle of the disk is 
$i \gsim 80^{\circ}$,
that is consistent with the geometrical configuration.

Similary, the canonical mass accretion rate is estimated to be
$\dot{M}/\dot{M}_{\rm HL} \lsim 0.8$
(or $\dot{M}/\dot{M}_{\rm E}=\Gamma \gsim 0.1$,
$\dot{M} \sim 10^{19}{\rm ~g~s}^{-1}$).
Since the mass ejection rate of jets is to be
$M_{\rm jet} = 10^{16}{\rm ~g~s}^{-1}$ (appendix),
a small fraction $(\sim 10^{-3})$ of accretion mass is ejected by jets.

\begin{flushleft}
{\bf 6. Concluding Remarks}
\end{flushleft}

We have examined Hoyle--Lyttleton accretion 
onto the central accretion disk.
Such a situation is very likely to be realized in the universe.
Because the radiation field of accretion disks is {\it anisotropic},
the accretion nature is quite different from
that of the spherical case,
qualitatively as well as quantitatively.
In contrast to the traditional spherical model,
the accretion radius $R_{\rm acc}$ generally depends on
an inclination angle $i$ between the accretion axis
and the symmetry axis of the disk
and the azimuthal angle $\varphi$ around the accretion axis,
as well as the mass $M$, 
the normalized luminosity $\Gamma$, 
and the relative velocity $v_\infty$.
Hence, the effective cross section of accretion changes,
and therefore, the accretion rate $\dot{M}$
depends on $i$ as well as $M$, $\Gamma$, and $v_\infty$.
In the case of a pole-on accretion ($i=0$),
the accretion radius becomes smaller than
that of the spherical case,
although the axisymmetry around $\varphi$ is retained.
The accretion rate is therefore smaller than
that of the spherical case.
In the case of an edge-on accretion ($i=90^{\circ}$),
the accretion radius depends strongly on $\varphi$
and is somewhat larger than that of the spherical case.
The accretion rate is larger than that of the spherical case.
Since accretion mainly takes place in the disk plane,
the anisotropic radiation fields
of the central accretion disks
automatically act to maintain the disk plane 
(the direction of jets associated with the disk is
also maintained).

We make some comments on recent numerical simulations.
In 2D non-axisymmetric simulations of black-hole accretion,
the so-called flip-flop instability occurs
and the accretion disk forms transiently
(Fryxell, Taam 1988; Taam, Fryxell 1989; Matsuda et al. 1991).
In 3D simulations, however,
the flip-flop instability disappears
and the accretion cone remains quite stable
(Ruffert, Arnett 1994; Ruffert 1994, 1996).
In 3D relativistic computations,
accretion also takes place in a stationary way without instability
(Font, Ib\'a\~nez 1998a, b).
According to these recent results
of 3D numerical simulations,
an accretion disk hardly forms via black-hole accretion,
even if transiently.
In some astronomical systems such as 1E~1740.7$-$2942,
on the other hand,
it is believed that
an accretion disk and jets exist
around an isolated black hole.
The present mechanism interprets the discrepancy 
between the numerical simulations and observational facts.
The accretion disk, once formed,
must break the axial symmetry of black-hole accretion!

We have neglected the effect of radiation drag.
If the radiation drag force is included,
the accretion radius will generally increase
(Nio et al. 1998).
When the relative velocity is large,
such radiation drag becomes important.
In the present study
we have assumed that
the size of the accretion disk is negligible
compared to the accretion radius.
This assumption is adequate for a slow accretor
[cf. equation (1)].
However, when the relative velocity is large,
the accretion radius could become as small as the disk.
In such a case, the size effect will be significant,
the radiation fields of the disk is complicated (Tajima, Fukue 1998),
and the present results would be modified quantitatively.
Magneto-hydrodynamical and relativistic effects are also ignored.
Finally,
the present mechanism can be applied
to the dust accretion onto isolated stars,
including protostars and late type stars.
These problems are left as future work.

\vspace{0.5cm}
\noindent
{\bf Reference}\re
Bondi H. 1952, MNRAS 112, 195 \re
Bondi H., Hoyle F. 1944, MNRAS 104, 273 \re
Bouchet, L. et al. 1991, ApJL 383, L45 \re
Font J.A., Ib\'a\~nez J.M. 1998a, ApJ 494, 297 \re
Font J.A., Ib\'a\~nez J.M. 1998b, MNRAS 298, 835 \re
Fryxell B.A., Taam R.E., McMillan S.L.W. 1987, ApJ 315, 536 \re
Fryxell B.A., Taam R.E. 1988, ApJ 335, 862 \re
Fukue J. 1986, PASJ 38, 167 \re
Fukue J., Kato S., Matsumoto R. 1985, PASJ 37, 383 \re
Ho C., Taam R.E., Fryxell B.A., Matsuda T., Koide H., Shima E.
   1989, MNRAS 238, 1447 \re
Hoyle F., Lyttleton R.A. 1939, Proc.Camb.Phil.Soc. 35, 405 \re
Hsieh S.-H., Spiegel E.A. 1976, ApJ 207, 244 \re
Hunt R. 1971, MNRAS 154, 141 \re
Johnson III W.N., Harnden F.R., Haymes R.C. 1972, ApJL 172, L1 \re
Johnson III W.N., Haymes R.C. 1973, ApJ 184, 103 \re
Kato S., Fukue J., Mineshige S. 1998,
   Black-Hole Accretion Disks (Kyoto University Press, Kyoto) 
   chap16 \re
Leventhal M., MacCallum C.J., Stang P.D. 1978, ApJL 225, L11 \re
Matsuda T. Sekino N., Sawada K., Shima E., Livio M., Anzer U.,
   B\"orner G. 1991, A\&A 248, 301 \re
Mirabel I.F., Morris M., Wink J., Paul J., Cordier B. 1991, 
   A\&A 251, L43 \re
Mirabel I.F., Rodriguez L.F., Cordier B., Paul J., Lebrun F. 1992, 
   Nature 358, 215 \re
Morris M. (ed) 1989, The Center of the Galaxy (Kluwer, Dordrecht) \re
Nio T., Matsuda T., Fukue J. 1998, PASJ 50, 495 \re
Ramaty R., Leventhal M., Chan K.W., Lingenfelter R.G. 1992, 
   ApJL 392, L63 \re
Ruffert M. 1994, ApJ 427, 342 \re
Ruffert M., Arnett D. 1994, ApJ 427, 351 \re
Ruffert M. 1996, A\&A 311, 817 \re
Shima E., Matsuda T., Takeda H., Sawada K. 1985, MNRAS 217, 367 \re
Sunyaev R. et al. 1991, A\&A 247, L29 \re
Taam R.E., Fryxell B.A. 1989, ApJ 339, 297 \re
Taam R.E., Fu A., Fryxell B.A. 1991, ApJ 371, 696 \re
Tajima Y., Fukue J. 1998, PASJ 50, 483 \re

\begin{flushleft}
{\bf Appendix. Physical Quantities of 1E~1740.7$-$2942}
\end{flushleft}

\renewcommand{\theequation}{A\arabic{equation}}
\setcounter{equation}{0}

In this appendix
we shall evaluate the physical quantities 
of the electron-positron pair jet,
i.e., the velocity and number density,
using the observational data and appropriate assumptions.

Let us first suppose that the observed radio jets from 
the X-ray source
1E~1740.7$-$2942 are composed of an electron-positron pair plasma
traveling with high velocity (Fukue 1986; Tajima, Fukue 1998).
Entrainment from the ambient gas must be negligible.
Or, the redshifted annihilation lines which are produced
in the entrainment region would be detected.
The electron-positron pairs are carried from the X-ray source
1E~1740.7$-$2942 to the annihilation point in the molecular cloud
without significant loss due to pair annihilation,
since the annihilation efficiency in the pair jet is low (Fukue 1986).
Thus, the mass is conserved along the jet:
\begin{equation}
     2\pi a^2 n_{\rm e}m_{\rm e}cu = \dot{M}_{\rm jet},
\label{A1}
\end{equation}
where $a$ is the radius of the pair jet, $n_{\rm e}$ 
the number density of electrons (which is equal to that of positrons),
$m_{\rm e}$ the mass of electrons,
$c$ the light speed, $u (=\gamma v/c)$ the four velocity 
of the pair jet, and $\dot{M}_{\rm jet}$ the mass ejection rate.
Of these quantities, the mass ejection rate $\dot{M}_{\rm jet}$
can be derived from the observed annihilation rate:
the annihilation rate of 
$10^{-3}$photons s${}^{-1}$ cm${}^{-2}$
corresponds to the mass ejection rate 
of $\sim 10^{16} {\rm g \, s}^{-1}$.
 From the VLA map (Mirabel et al. 1992)
the radius $a$ of the pair jet is about 0.1--0.2pc.
Hence, equation (\ref{A1}) gives one relation 
between $n_{\rm e}$ and $u$.


At the distance of $\sim 1$~pc from 
the central X-ray source 1E~1740.7$-$2942,
the pair jet terminates and electrons and positrons are 
annihilated there (Mirabel et al. 1991).
The momentum carried by the pair jet should be balanced with
that of the interstellar gas.
The former is 
$2(n_{\rm e}m_{\rm e}c^2 + n_{\rm e}kT_{\rm e})u^2 + 2n_{\rm e}
kT_{\rm e}$, where $k$ is the Boltmann constant and 
$T_{\rm e}$ the electron temperature of the pair jet.
The factor 2 includes the contribution from both electrons and positrons.
Since it is reasonable that the pair jet is sufficiently cold
(Fukue 1986),
this reduces to $2n_{\rm e}m_{\rm e}c^2u^2$,
which just expresses the ram pressure of the pair jet.
On the other hand, the interstellar gas pressure is $nkT$, where
$n$ is the number density of the interstellar gas in the molecular
cloud and $T$ is its temperature.
Then, the condition that the ram pressure of the pair jet 
is balanced with the gas pressure of the ambient molecular gas becomes
\begin{equation}
     2n_{\rm e}m_{\rm e}c^2u^2 = nkT.
\label{A2}
\end{equation}
In this equation (\ref{A2}), 
the number density $n$ of the interstellar gas 
in the molecular cloud is of the order of 
$5\times 10^4 {\rm ~cm}^{-3}$ 
(Mirabel et al. 1991).
The temperature $T$ of the molecular gas is typically 10--20~K.
Hence, equation (\ref{A2}) gives another relation 
between $n_{\rm e}$ and $u$.

Solving equations (\ref{A1}) and (\ref{A2}), we have typically
\begin{eqnarray}
     u & = & \pi a^2 nkT/\dot{M}_{\rm jet} c \nonumber \\
       & = & 0.275 \left(\frac{a}{\rm 0.2pc}\right)^2 
             \frac{n}{5 \times 10^4 {\rm cm}^{^3}} 
             \frac{T}{10{\rm K}} 
  \left(\frac{\dot{M}_{\rm jet}}{10^{16}{\rm g \, s}^{-1}}\right)^{-1},
\label{A3}
\end{eqnarray}
\begin{eqnarray}
     n_{\rm e} & = & \dot{M}_{\rm jet}^2/2\pi^2 a^4 m_{\rm e} nkT 
\nonumber \\
      & = & 5.55 \times 10^{-4} 
      \left(\frac{a}{\rm 0.2pc}\right)^{-4} 
      \left(\frac{n}{5 \times 10^4 {\rm cm}^{^3}}\right)^{-1} 
\nonumber \\
      &&\times
      \left(\frac{T}{10{\rm K}}\right)^{-1} 
      \left(\frac{\dot{M}_{\rm jet}}{10^{16}{\rm g \, s}^{-1}}\right)^2 
     {\rm ~cm}^{-3}.
\end{eqnarray}

The velocity of the electron-positron pair jet from the X-ray source
1E~1740.7$-$2942 is evaluated to be mildly relativistic.
It is emphasized that
the typical velocity given by equation (\ref{A3}) 
is close to that of SS433 (0.26$c$).

\newpage

\begin{center}
Figure Captions
\end{center}

Fig. 1.
Configuration of the system.
A point object with mass $M$,
which is surrounded by an accretion disk with luminosity $L$,
is moving against an ambient gas of density $\rho_\infty$
at relative speed $v_\infty$.
The accretion disk is generally inclined to the gas flow
with inclination angle $i$.
Left: the side view from the $y$-direction.
Right: the top view from the $z$-direction.

Fig. 2.
Accretion radius normalized by the Hoyle--Lyttleton one,
$R_{\rm acc}/R_{\rm HL}$,
as a function of the normalized luminosity $\Gamma$.
The dotted line, 
which linearly decreases with $\Gamma$,
is for the spherical case.
The dashed curve
is for the case of pole-on accretion ($i=0$).
The solid curves
are for the case of edge-on accretion ($i=90^{\circ}$),
where $\varphi=90^{\circ}$ to 0 from top to bottom.

Fig.3.
Shapes of accretion cross section in the case of edge-on accretion
($i=90^{\circ}$) for various $\Gamma$.
The outer circles represent the classical Hoyle--Lyttleton radii,
while the inner circles denote those of the spherical case.

Fig.4.
Cross sectional area
normalized by the Hoyle--Lyttleton value,
 $A_{\rm acc}/\pi R_{\rm HL}^2$,
as a function of the normalized luminosity $\Gamma$.
The dotted curve is for the spherical case.
The dashed curve
is for the case of pole-on accretion ($i=0$).
The solid curve
is for the case of edge-on accretion ($i=90^{\circ}$).
This plot also represents
the mass accretion rate
normalized by the Hoyle--Lyttleton one,
$\dot{M}/\dot{M}_{\rm HL}$.

Fig.5.
Canonical luminosity
normalized by the Eddington one,
$\Gamma_{\rm can}$ ($=L_{\rm can}/L_{\rm E}$),
as a function of $\dot{m}_{\rm HL}$
($=\dot{M}_{\rm HL}/\dot{M}_{\rm E}$).
The dotted curve is for the spherical case.
The dashed curve
is for the case of pole-on accretion ($i=0$).
The solid curve
is for the case of edge-on accretion ($i=90^{\circ}$).

\end {document}